\documentstyle[aps,prl]{revtex}
\draft

\begin{document}
\author{Di Qing, Xiang-Song Chen, and Fan Wang}
\address{Department of Physics and Center for Theoretical Physics, Nanjing
University, Nanjing 210093, People's Republic of China}
\title{Is Nucleon Spin Structure Inconsistent with Constituent Quark Model?}
\maketitle

\begin{abstract}
A qualitative QCD analysis and a quantitative model calculation are given to
show that the constituent quark model remains a good approximation even
with the nucleon spin structure revealed in polarized deep inelastic 
scattering taking into account.
\end{abstract}

\pacs{PACS numbers: 14.20.Dh, 12.39.Jh, 24.85.+p, 13.88+e}

\section{Introduction}

Hadron structure studies might be traced back to Fermi-Yang \cite{a1} and
Sakata models \cite{a2}. Gell-Mann \cite{a3} and Zweig \cite{a4} proposed
the quark model of hadrons. Lepton-nucleon deep inelastic scattering (DIS) 
\cite{a5} verified the quark structure of hadrons. However the quark
revealed in DIS is found to be different from the quark as a carrier of
SU(3) symmetry in Gell-Mann-Zweig model. The former is almost a free 
particle, the later is strongly bounded. Even though this lead Feynman 
\cite{a6} to call the quark detected in DIS as parton, such qualitatively
different behavior of quark didn't hurt the hadron structure studies in both
directions. On the contrary, phenomenological success of SU(6) quark model
in explaining the hadron properties and the evidence obtained in DIS for the
existence of quark inside hadrons worked together and motivated the
development of a new strong interaction theory, the quantum chromodynamics
(QCD) \cite{a7}. The asymptotic freedom and confinement properties of QCD
fitted perfectly weak interaction parton picture revealed in DIS and the
fact that no free quark was discovered in all intensive experimental
searches. The weak interacting high energy process can be calculated and
tested due to asymptotic freedom of QCD and gave strong support to this new
strong interaction theory. However the hadron structure and low energy
hadron interactions are hard to be calculated due to confinement. Lattice QCD
is promising to find low energy solution but still suffers numerical
uncertainty for the time being. Various QCD models developed under this
condition. Different models emphasize different effective degree of freedom
inspired by QCD properties \cite{a8}. Among them, constituent quark model is
the most successful one in explaining hadron properties \cite{a9} and hadron
interactions \cite{a10}; and gives the most popular intuitive picture of
hadron internal structure.

The most striking feature of constituent quark model is that it gives a very
simple but quite successful explanation of the baryon spin and magnetic
moment by means of effective constituent quark masses. Once again, one meets
the qualitatively different behaviors of quark, i.e., the constituent quark
mass needed in the hadron spectroscopy is much larger than the current quark
mass revealed in high energy processes. This lead Weinberg to ask ``why do
quarks behave like bare Dirac particles'' \cite{a11}; and the relation
between constituent quark and current quark is a holy grail in hadron
physics. In the $\left( 1s\right) ^3$ pure valence nonrelativistic
constituent quark model, the nucleon spin is solely carried by quark spin,
the orbital angular momentum is zero because quarks are assumed to be in the
lowest $s$-wave $\left( 1s\right) $ state. The nucleon magnetic moment is
also solely contributed by quark spin magnetic moments. In 1988, EMC group 
\cite{a12} measured the polarization asymmetry of polarized $\mu $-proton
deep inelastic scattering and extracted the proton spin structure function which
showed that quark spin contributes only a small amount of the proton spin.
Constituent quark model has been challenged by this surprising result and 
lead to the proton spin crisis. Many models and mechanisms have
been invoked to explain why quark spin contribution is suppressed and how to
supply angular momentum to compensate the missing spin of nucleon \cite{a13}%
. After ten years intensive studies both experimentally and theoretically,
the prevailing view point seems to be that the nucleon spin structure
discovered in DIS is inconsistent with constituent quark model. Only a
minority \cite{a14} keeps the view point that quark spin is primarily
responsible for generating the nucleon spin because in the valence quark
region the polarization asymmetry confirmed the constituent quark model
prediction. The sea quark component neglected as an approximation in pure
valence quark model plays a vital role in suppressing the quark spin
contribution $\Delta q$ extracted from DIS.

This report stands by the minority through both a qualitative QCD analysis
and a quantitative model calculation. In section II, the difference between
the quark spin sum $\Delta \Sigma $ of the constituent quark model and the $%
\Delta q$ measured in DIS is explained. In section III, the duality of
nucleon spin structure is explained. In section IV, QCD relations of
baryon spin, magnetic moment, and tensor charge are derived and
discussed. A constituent quark model with valence-sea quark component mixing is
shown to be able to reconcile the difference between the quark spin sum $%
\Delta \Sigma $ and $\Delta q$ and fit other baryon properties as well in
section V. The discussions and conclusions are put in section VI.

\section{The difference between the quark spin sum $\Delta \Sigma $ of the
constituent quark model and the quark spin contribution $\Delta \lowercase{q}$ measured
in DIS}

For a constituent quark model, the quark spin sum of proton (in general
baryon) at rest can be expressed as 
\begin{eqnarray}
\Delta \Sigma &=&\sum_i\int d^3k\left( q_{\uparrow }^i\left( \stackrel{%
\rightarrow }{k}\right) -q_{\downarrow }^i\left( \stackrel{\rightarrow }{k}%
\right) +\overline{q}_{\uparrow }^i\left( \stackrel{\rightarrow }{k}\right) -%
\overline{q}_{\downarrow }^i\left( \stackrel{\rightarrow }{k}\right) \right)
\nonumber   \\
&=&\sum_i\left( q_{\uparrow }^i-q_{\downarrow }^i+\overline{q}_{\uparrow }^i-%
\overline{q}_{\downarrow }^i\right),
\end{eqnarray}
here$\ q_{\uparrow \downarrow }^i$ ($\overline{q}_{\uparrow \downarrow }^i$)
means quark (antiquark) of flavor $i$ with spin parallel or antiparallel to
the baryon spin. $\stackrel{\rightarrow }{k}$ is used to show the momentum
distribution.

The quark spin contribution $\Delta q$ measured in DIS is defined as 
\begin{equation}
\left\langle PS\left| \int d^3x\overline{\psi }\gamma ^\mu \gamma ^5\psi
\right| PS\right\rangle =S_\mu \cdot \Delta q,  
\end{equation}
where $S_\mu $ is the proton polarization vector, $\int d^3x\overline{\psi }%
\gamma ^\mu \gamma ^5\psi $ is the quark axial vector current. For the
parton model in the infinite momentum frame 
\begin{equation}
\Delta q=\sum_i\int dx\left( q_{\uparrow }^i\left( x\right) -q_{\downarrow
}^i\left( x\right) +\overline{q}_{\uparrow }^i\left( x\right) -\overline{q}%
_{\downarrow }^i\left( x\right) \right),
\end{equation}
where $q_{\uparrow \downarrow }^i\left( x\right) $ ($\overline{q}_{\uparrow
\downarrow }^i\left( x\right) $) is the probability of finding a quark
(antiquark) with fraction $x$ of the proton momentum and polarization parallel
or antiparallel to the proton spin. Even though the expressions (1) and (3)
are similar, the physical meaning is not the same in general. To show the
difference, let's express the quark axial vector current operator in
terms of Pauli spin 
\begin{eqnarray}
\int d^3x\overline{\psi }\stackrel{\rightarrow }{\gamma }\gamma ^5\psi
&=&\sum_{i\lambda \lambda ^{/}}\int d^3k\chi _\lambda ^{\dagger }\stackrel{%
\rightarrow }{\sigma }\chi _{\lambda ^{^{\prime }}}\left( a_{i\stackrel{%
\rightarrow }{k}\lambda }^{\dagger }a_{i\stackrel{\rightarrow }{k}\lambda
^{^{\prime }}}-b_{i\stackrel{\rightarrow }{k}\lambda ^{^{\prime }}}^{\dagger
}b_{i\stackrel{\rightarrow }{k}\lambda }\right)  \nonumber \\
&&\ \ \ -\sum_{i\lambda \lambda ^{/}}\int d^3k\chi _\lambda ^{\dagger }\frac{%
\stackrel{\rightarrow }{\sigma }\cdot \stackrel{\rightarrow }{k}}{k_0\left(
k_0+m_i\right) }i\stackrel{\rightarrow }{\sigma }\times \stackrel{\rightarrow 
}{k}\chi _{\lambda ^{^{\prime }}}\left( a_{i\stackrel{\rightarrow }{k}%
\lambda }^{\dagger }a_{\stackrel{\rightarrow }{k}\lambda ^{^{\prime }}}-b_{i%
\stackrel{\rightarrow }{k}\lambda ^{^{\prime }}}^{\dagger }b_{i\stackrel{%
\rightarrow }{k}\lambda }\right)  \nonumber \\
&&\ \ \ +\sum_{i\lambda \lambda ^{/}}\int d^3k\chi _\lambda ^{\dagger }\frac{%
i\stackrel{\rightarrow }{\sigma }\times \stackrel{\rightarrow }{k}}{k_0}\chi
_{\lambda ^{^{\prime }}}a_{i\stackrel{\rightarrow }{k}\lambda }^{\dagger
}b_{i-\stackrel{\rightarrow }{k}\lambda ^{^{\prime }}}^{\dagger }+h.c.
\end{eqnarray}
In getting this expression an expansion 
\begin{equation}
\psi _i\left( x\right) =\left( 2\pi \right) ^{-\frac 32}\sum_\lambda \int
d^3k\left( a_{_i\stackrel{\rightarrow }{k}\lambda }u_{\stackrel{\rightarrow 
}{k}\lambda }e^{i\stackrel{\rightarrow }{k}\cdot \stackrel{\rightarrow }{x}%
}+b_{i\stackrel{\rightarrow }{k}\lambda }^{\dagger }v_{\stackrel{\rightarrow 
}{k}\lambda }e^{-i\stackrel{\rightarrow }{k}\cdot \stackrel{\rightarrow }{x}%
}\right) 
\end{equation}
has been used. Here $a_{\stackrel{\rightarrow }{k}\lambda }^{\dagger }$ ($b_{%
\stackrel{\rightarrow }{k}\lambda }^{\dagger }$) is quark (antiquark) of
flavor $i$ creation operator with momentum $\stackrel{\rightarrow }{k}$ and
polarization $\lambda $ in Heisenberg representation. $u_{\stackrel{%
\rightarrow }{k}\lambda }$, $v_{\stackrel{\rightarrow }{k}\lambda }$ are
usual Dirac spinors, $\chi _{\lambda}$ is Pauli spinor. Flavor
spinor wavefunction is omitted in Eq.(5). $k_0$ ($\stackrel{\rightarrow }{k}$%
) and $m_i$ are energy (momentum) and mass of quarks. We can't identify the $%
a_{\stackrel{\rightarrow }{k}\lambda }^{\dagger }$, $b_{\stackrel{%
\rightarrow }{k}\lambda }^{\dagger }$ with the constituent quark and
antiquark creation operators. However Eq.(4) shows at least that for any
realistic proton state, the matrix element of the quark axial vector current
operator, usually called the quark spin contribution to the nucleon spin, is
not solely due to Pauli spin contribution (the first term in Eq.(4)), but
also a contribution from quark orbital motion (the second term in Eq.(4)).
Only in special cases, such as for static quark model (all quark momentum $%
\stackrel{\rightarrow }{k}$ is assumed to be zero), the second terms does
not contribute. Another case is the parton model in the infinite momentum
frame where $k_{\perp }/k_0$ is negligible and therefore the second term
does not contribute either. However it should be noted that it is the matrix
element of the axial vector current operator evolved to be the helicity difference
in Eq.(3) which should be compared to the matrix element of the whole axial
vector current operator in Eq.(4) for a proton at rest rather than the
matrix element of the first term in Eq.(4). The Pauli spin itself is not a
Lorentz invariant quantity as had been pointed out by Ma \cite{a15}.

In addition, pure valence quark configuration is an approximation. Sea quark
component should be considered even for a constituent quark model from a
general view point of Fock space expansion. After including the valence and
sea quark components mixing in the model Fock space, the third term of Eq.(4)
(pair
creation and annihilation term) will also contribute to the $\Delta q$
defined in Eq.(2). This would make the quark spin sum $\Delta \Sigma $
deviating from the quark spin contribution $\Delta q$ further. This effect
has rarely been calculated in quark model approach \cite{a16}. It will be
shown in a valence-sea quark mixing model calculation (see section V) that
this is an important correction to $\Delta q$ even for a $15\%$ sea quark
component mixing.

Neglecting the antiquark and the pair creation (annihilation) term and 
assuming the quark moment distribution of nucleon ground state to be
spherically symmetric, Eq.(4) reduces to the Melosh rotation result
discussed by Ma\cite{a15}.

To sum up, the quark spin sum $\Delta \Sigma$ of a constituent quark model
should not be identified with the quark spin contribution $\Delta q$
measured in DIS. The fact that $\Delta q$ discovered in DIS is much smaller (see next section)
than the constituent quark spin sum $\Delta \Sigma $ has not proved yet that the
nucleon spin structure is inconsistent with the constituent quark model.

\section{Duality of nucleon spin structure}

After ten years intensive theoretical and experimental studies, now the
world average of quark spin contribution to the proton spin is \cite{a17} 
\begin{eqnarray}
\Delta q\left( 3~GeV^2\right) &=&\Delta u+\Delta d+\Delta s  \nonumber \\
\ &=&0.82(1)-0.44(1)-0.11(1)  \nonumber  \\
\ &=&0.27(4).
\end{eqnarray}
We have explained in section II that there is not really a serious contradiction
between the DIS measurement and the constituent quark model picture, even
though for example, a pure valence nonrelativistic constituent quark model
gives 
\begin{equation}
\Delta u=\frac 43,\Delta d=-\frac 13,\Delta s=0,\Delta \Sigma =1.
\end{equation}
Because $\Delta q$ and $\Delta \Sigma$ are not the same matrix element
of proton.

However one question should be answered that where does the proton get
additional angular momentum if one follows the QCD view point where quark
axial vector current is only part of the source of the proton spin. Proton,
as a QCD system, its angular momentum is in general consisted of\cite{a18} 
\begin{eqnarray}
\stackrel{\rightarrow }{J}_{QCD} &=&\stackrel{\rightarrow }{S}_q+\stackrel{%
\rightarrow }{L}_q+\stackrel{\rightarrow }{S}_g+\stackrel{\rightarrow }{L}_g
\nonumber \\
\ &=&\frac 12\int d^3x\psi ^{\dagger }\stackrel{\rightarrow }{\Sigma }\psi
+\int d^3x\psi ^{\dagger }\stackrel{\rightarrow }{x}\times \frac 1i\stackrel{%
\rightarrow }{\partial }\psi  \nonumber \\
&&\ \ +\int d^3x\stackrel{\rightarrow }{E}\times \stackrel{\rightarrow }{A}%
+\int d^3xE_i\stackrel{\rightarrow }{r}\times \stackrel{\rightarrow }{%
\partial }A_i. 
\end{eqnarray}

In a constituent quark model, the explicit gluon degree of freedom is
usually neglected and its effect is included in the quark interaction term.
Therefore only quark orbital angular momentum $\stackrel{\rightarrow }{L}_q$
will be able to contribute additional angular momentum to proton spin
besides the quark axial vector current operator $\stackrel{\rightarrow }{S}%
_q $ in the constituent quark model space. Let's also express the quark
orbital angular momentum $\stackrel{\rightarrow }{L}_q$ in terms of Pauli
spinor, 
\begin{eqnarray}
\stackrel{\rightarrow }{L}_q &=&\sum_{i\lambda }\int d^3k\left( a_{i%
\stackrel{\rightarrow }{k}\lambda }^{\dagger }i\stackrel{\rightarrow }{%
\partial }_k\times \stackrel{\rightarrow }{k}a_{i\stackrel{\rightarrow }{k}%
\lambda }+b_{i\stackrel{\rightarrow }{k}\lambda }^{\dagger }i\stackrel{%
\rightarrow }{\partial }_k\times \stackrel{\rightarrow }{k}b_{i\stackrel{%
\rightarrow }{k}\lambda }\right)  \nonumber \\
&&\ +\frac 12\sum_{\lambda \lambda ^{/}}\int d^3k\chi _\lambda ^{\dagger }%
\frac{\stackrel{\rightarrow }{\sigma }\cdot \stackrel{\rightarrow }{k}}{%
k_0\left( k_0+m\right) }i\stackrel{\rightarrow }{\sigma }\times \stackrel{%
\rightarrow }{k}\chi _{\lambda ^{^{\prime }}}\left( a_{i\stackrel{%
\rightarrow }{k}\lambda }^{\dagger }a_{i\stackrel{\rightarrow }{k}\lambda
^{^{\prime }}}-b_{i\stackrel{\rightarrow }{k}\lambda ^{^{\prime }}}^{\dagger
}b_{i\stackrel{\rightarrow }{k}\lambda }\right)  \nonumber \\
&&\ -\sum_{i\lambda \lambda ^{/}}\int d^3k\chi _\lambda ^{\dagger }\frac{i%
\stackrel{\rightarrow }{\sigma }\times \stackrel{\rightarrow }{k}}{2k_0}\chi
_{\lambda ^{^{\prime }}}a_{i\stackrel{\rightarrow }{k}\lambda }^{\dagger
}b_{i-\stackrel{\rightarrow }{k}\lambda ^{^{\prime }}}^{\dagger }+h.c..
\end{eqnarray}
It is interesting to note that the second and third term in Eq.(4) and (9)
cancel with each other exactly. We have 
\begin{eqnarray}
\stackrel{\rightarrow }{S}_q+\stackrel{\rightarrow }{L}_q &=&\stackrel{%
\rightarrow }{S}_q^{NR}+\stackrel{\rightarrow }{L}_q^{NR}  \nonumber
 \\
\ &=&\frac 12\sum_{i\lambda \lambda ^{/}}\int d^3k\chi _\lambda ^{\dagger }%
\stackrel{\rightarrow }{\sigma }\chi _{\lambda ^{^{\prime }}}\left( a_{i%
\stackrel{\rightarrow }{k}\lambda }^{\dagger }a_{i\stackrel{\rightarrow }{k}%
\lambda ^{^{\prime }}}-b_{i\stackrel{\rightarrow }{k}\lambda ^{^{\prime
}}}^{\dagger }b_{i\stackrel{\rightarrow }{k}\lambda }\right)  \nonumber \\
&&\ \ +\sum_{i\lambda }\int d^3k\left( a_{i\stackrel{\rightarrow }{k}\lambda
}^{\dagger }i\stackrel{\rightarrow }{\partial }_k\times \stackrel{%
\rightarrow }{k}a_{i\stackrel{\rightarrow }{k}\lambda }+b_{i\stackrel{%
\rightarrow }{k}\lambda }^{\dagger }i\stackrel{\rightarrow }{\partial }%
_k\times \stackrel{\rightarrow }{k}b_{i\stackrel{\rightarrow }{k}\lambda
}\right) .
\end{eqnarray}
where $\stackrel{\rightarrow }{S}_q^{NR}$ $\stackrel{\rightarrow }{L}_q^{NR}$
represent the first term in Eq.(4) and (9). We call them nonrelativistic
quark spin and nonrelativistic quark orbital angular momentum because they
can be related to the constituent quark model spin and orbital angular
momentum if we make an assumption that the quark (antiquark) creation
operator $a_{ks}^{\dagger }$ ($b_{ks}^{\dagger }$) can be directly related
to the effective constituent quark degree of freedom (see section V).
Eq.(10) tells us that there are two equivalent decompositions of the proton
spin, either in terms of the relativistic $\stackrel{\rightarrow }{S}_q$ and 
$\stackrel{\rightarrow }{L}_q$ directly derived from QCD Lagrangian, or in
terms of the nonrelativistic quark spin $\stackrel{\rightarrow }{S}_q^{NR}$
and orbital angular momentum $\stackrel{\rightarrow }{L}_q^{NR}$. Under the
assumption that the Heisenberg operators used in Eq.(5) can be related to
the constituent quark degree of freedom, then in a pure $s$-wave
nonrelativistic constituent quark model we would have the picture that the
nucleon spin can either attributed solely to constituent quark spin $%
\stackrel{\rightarrow }{S}_q^{NR}$ and orbital angular momentum $\stackrel{%
\rightarrow }{L}_q^{NR}$ would not contribute (we have this picture already
more than twenty years); or attributed to the relativistic quark spin $%
\stackrel{\rightarrow }{S}_q$ which is reduced due to the orbital motion of
quarks and the sea quark pair creation (annihilation) process (see section
V), and the relativistic angular momentum $\stackrel{\rightarrow }{L}_q$
will contribute compensation terms to make the total proton spin unchanged
even for a pure $s$-wave constituent quark model state. We call this the 
duality of nucleon spin structure. The above discussion
is obviously also true for other baryons.

\section{QCD relations among baryon spin, magnetic moment, and tensor charge}

Now let's turn to baryon magnetic moment 
\begin{eqnarray}
\mu _B &=&\left\langle B\left| \left( \stackrel{\rightarrow }{\mu }_B\right)
_3\right| B\right\rangle  \nonumber  \\
\ &=&\left\langle B\left| \sum_i\frac{Q_i}2\int d^3x\psi _i^{\dagger }\left( 
\stackrel{\rightarrow }{x}\times \stackrel{\rightarrow }{\alpha }\right)
_3\psi _i\right| B\right\rangle .
\end{eqnarray}
Use the same Fourier expansion (5), we obtain 
\begin{eqnarray}
\stackrel{\rightarrow }{\mu }_B &=&\sum_i\sum_\lambda \int d^3k\frac{Q_i}{%
2k_0}\left( a_{i\stackrel{\rightarrow }{k}\lambda }^{\dagger }i\stackrel{%
\rightarrow }{\partial }_k\times \stackrel{\rightarrow }{k}a_{i\stackrel{%
\rightarrow }{k}\lambda }-b_{i\stackrel{\rightarrow }{k}\lambda }^{\dagger }i%
\stackrel{\rightarrow }{\partial }_k\times \stackrel{\rightarrow }{k}b_{i%
\stackrel{\rightarrow }{k}\lambda }\right)  \nonumber \\
&&\ \ +\sum_i\sum_{\lambda \lambda ^{/}}\int d^3k\frac{Q_i}{2k_0}\chi
_\lambda ^{\dagger }\stackrel{\rightarrow }{\sigma }\chi _{\lambda
^{^{\prime }}}\left( a_{i\stackrel{\rightarrow }{k}\lambda }^{\dagger }a_{i%
\stackrel{\rightarrow }{k}\lambda ^{^{\prime }}}+b_{i\stackrel{\rightarrow }{%
k}\lambda ^{^{\prime }}}^{\dagger }b_{i\stackrel{\rightarrow }{k}\lambda
}\right)  \nonumber \\
&&\ \ -\sum_i\sum_{\lambda \lambda ^{/}}\int d^3k\frac{Q_i}{2k_0}\chi
_\lambda ^{\dagger }\frac{\stackrel{\rightarrow }{\sigma }\cdot \stackrel{%
\rightarrow }{k}}{2k_0\left( k_0+m_i\right) }i\stackrel{\rightarrow }{\sigma 
}\times \stackrel{\rightarrow }{k}\chi _{\lambda ^{^{\prime }}}\left( a_{i%
\stackrel{\rightarrow }{k}\lambda }^{\dagger }a_{i\stackrel{\rightarrow }{k}%
\lambda ^{^{\prime }}}+b_{i\stackrel{\rightarrow }{k}\lambda ^{^{\prime
}}}^{\dagger }b_{i\stackrel{\rightarrow }{k}\lambda }\right)  \nonumber \\
&&\ \ -\sum_i\sum_\lambda \int d^3k\frac{Q_i}{2k_0}\frac{\stackrel{%
\rightarrow }{k}}{2\left( k_0+m_i\right) }a_{i\stackrel{\rightarrow }{k}%
\lambda }^{\dagger }b_{i-\stackrel{\rightarrow }{k}\lambda }^{\dagger}+h.c.
  \nonumber
\\
&&\ \ +\sum_i\sum_{\lambda \lambda ^{/}}\int d^3k\frac{Q_i}{2k_0}a_{i%
\stackrel{\rightarrow }{k}\lambda }^{\dagger }i\stackrel{\rightarrow }{%
\partial }_kb_{i-\stackrel{\rightarrow }{k}\lambda ^{^{\prime }}}^{\dagger
}\times \chi _\lambda ^{\dagger }\left( m_i\stackrel{\rightarrow }{\sigma }+%
\frac{\stackrel{\rightarrow }{\sigma }\cdot \stackrel{\rightarrow }{k}}{%
k_0+m_i}i\stackrel{\rightarrow }{\sigma }\times \stackrel{\rightarrow }{k}%
\right) \chi _{\lambda ^{^{\prime }}}+h.c..
\end{eqnarray}
Use the expression (4) and (9) for $\stackrel{\rightarrow }{S}_q$ and $%
\stackrel{\rightarrow }{L}_q$, Eq.(12) can be reexpressed as 
\begin{eqnarray}
\stackrel{\rightarrow }{\mu }_B &=&\sum_i\int d^3k\frac{Q_i}{k_0}\left( 
\stackrel{\rightarrow }{S}_{i\stackrel{\rightarrow }{k}}-\stackrel{%
\rightarrow }{S}_{\overline{i}\stackrel{\rightarrow }{k}}\right) +\int d^3k%
\frac{Q_q}{2k_0}\left( \stackrel{\rightarrow }{L}_{i\stackrel{\rightarrow }{k%
}}-\stackrel{\rightarrow }{L}_{\overline{i}\stackrel{\rightarrow }{k}}\right)
\nonumber   \\
&&\ \ \ +\text{{\rm pair creation and annihination terms.}}
\end{eqnarray}
Here 
\begin{eqnarray}
\stackrel{\rightarrow }{S}_{i\stackrel{\rightarrow }{k}} &=&\frac 12%
\sum_{\lambda \lambda ^{/}}\chi _\lambda ^{\dagger }\left( \stackrel{%
\rightarrow }{\sigma }-\frac{\stackrel{\rightarrow }{\sigma }\cdot \stackrel{%
\rightarrow }{k}}{k_0\left( k_0+m_i\right) }i\stackrel{\rightarrow }{\sigma }%
\times \stackrel{\rightarrow }{k}\right) \chi _{\lambda ^{^{\prime }}}a_{i%
\stackrel{\rightarrow }{k}\lambda }^{\dagger }a_{i\stackrel{\rightarrow }{k}%
\lambda ^{^{\prime }}}  \nonumber \\
\stackrel{\rightarrow }{S}_{\overline{i}\stackrel{\rightarrow }{k}} &=&-%
\frac 12\sum_{\lambda \lambda ^{/}}\chi _\lambda ^{\dagger }\left( \stackrel{%
\rightarrow }{\sigma }-\frac{\stackrel{\rightarrow }{\sigma }\cdot \stackrel{%
\rightarrow }{k}}{k_0\left( k_0+m_i\right) }i\stackrel{\rightarrow }{\sigma }%
\times \stackrel{\rightarrow }{k}\right) \chi _{\lambda ^{^{\prime }}}b_{i%
\stackrel{\rightarrow }{k}\lambda ^{^{\prime }}}^{\dagger }b_{i\stackrel{%
\rightarrow }{k}\lambda }  \nonumber \\
\stackrel{\rightarrow }{L}_{i\stackrel{\rightarrow }{k}} &=&\sum_\lambda a_{i%
\stackrel{\rightarrow }{k}\lambda }^{\dagger }i\stackrel{\rightarrow }{%
\partial }_k\times \stackrel{\rightarrow }{k}a_{i\stackrel{\rightarrow }{k}%
\lambda }+\sum_{\lambda \lambda ^{/}}\chi _\lambda ^{\dagger }\frac{%
\stackrel{\rightarrow }{\sigma }\cdot \stackrel{\rightarrow }{k}}{2k_0\left(
k_0+m\right) }i\stackrel{\rightarrow }{\sigma }\times \stackrel{\rightarrow 
}{k}\chi _{\lambda ^{^{\prime }}}a_{i\stackrel{\rightarrow }{k}\lambda
}^{\dagger }a_{i\stackrel{\rightarrow }{k}\lambda ^{^{\prime }}}  \nonumber
\\
\stackrel{\rightarrow }{L}_{\overline{i}\stackrel{\rightarrow }{k}}
&=&\sum_\lambda b_{i\stackrel{\rightarrow }{k}\lambda ^{^{\prime
}}}^{\dagger }i\stackrel{\rightarrow }{\partial }_k\times \stackrel{%
\rightarrow }{k}b_{i\stackrel{\rightarrow }{k}\lambda }-\sum_{\lambda
\lambda ^{/}}\chi _\lambda ^{\dagger }\frac{\stackrel{\rightarrow }{\sigma }%
\cdot \stackrel{\rightarrow }{k}}{2k_0\left( k_0+m\right) }i\stackrel{%
\rightarrow }{\sigma }\times \stackrel{\rightarrow }{k}\chi _{\lambda
^{^{\prime }}}b_{i\stackrel{\rightarrow }{k}\lambda ^{^{\prime }}}^{\dagger
}b_{i\stackrel{\rightarrow }{k}\lambda }
\end{eqnarray}
If we restrict our discussion on the baryon ground states, it is plausible
to approximate the $k_0$ by its average $\left\langle k_0\right\rangle $
and take $\left\langle k_0\right\rangle $ as effective quark mass $m_i^{eff}$%
, we would have 
\begin{eqnarray}
\stackrel{\rightarrow }{\mu }_B&=&\sum_i\frac{Q_i}{m_i^{eff}}\left( \stackrel{%
\rightarrow }{S}_i-\stackrel{\rightarrow }{S}_{\overline{i}}\right) +\sum_i%
\frac{Q_i}{2m_i^{eff}}\left( \stackrel{\rightarrow }{L}_i-\stackrel{%
\rightarrow }{L}_{\overline{i}}\right)\nonumber\\
&&+{\rm pair~creation~and~annihilation~terms}, 
\end{eqnarray}
where $\stackrel{\rightarrow }{S}_{i(\overline{i})}=\int d^3k\stackrel{%
\rightarrow }{S}_{i(\overline{i})\vec{k}}$, $\stackrel{\rightarrow }{L}_{i(%
\overline{i})}=\int d^3k\stackrel{\rightarrow }{L}_{i(\overline{i})\vec{k}}$.
Eq.(15), after neglecting the pair terms, is quite similar to the magnetic 
moment operator that one has used
in the constituent quark model calculation. Suppose a baryon ground state
in Eq.(11) can be expressed by a Fock state of quark degree of freedom
(see Eq.(28)), the average is over all Fock components and all quark energy.
This average quark energy $%
\left\langle k_0\right\rangle $ bounded in a baryon plays the role of the
constituent quark mass, which explains why the constituent quark mass is much 
heavier than the current quark mass. However we have discussed before 
that it is not the
relativistic spin $\stackrel{\rightarrow }{S}_q$ and orbital angular
momentum $\stackrel{\rightarrow }{L}_q$ but the nonrelativistic $\stackrel{%
\rightarrow }{S}_q^{NR}$ and $\stackrel{\rightarrow }{L}_q^{NR}$ which are
closely related to the operators used in constituent quark model
calculations. In order to obtain the relation between baryon magnetic
moments and the nonrelativistic spin $\stackrel{\rightarrow }{S}_q^{NR}$ and
orbital angular momentum $\stackrel{\rightarrow }{L}_q^{NR}$, we need
another operator --- the tensor operator $\int d^3x\overline{\psi }\sigma
_{\mu \nu }\psi $ which was emphasized recently by MIT group \cite{a19}. 
Its matrix
element in a proton state is 
\begin{equation}
\left\langle PS\left| \int d^3x\overline{\psi }\sigma _{\mu \nu }\psi
\right| PS\right\rangle =\delta q\overline{u}\left( PS\right) \sigma _{\mu
\nu }u\left( PS\right),
\end{equation}
where $u\left( PS\right) $ is the Dirac spinor of proton and $\sigma _{\mu
\nu }=\frac i2\left( \gamma _\mu \gamma _\nu -\gamma _\nu \gamma _\mu
\right) $, $\delta q$ is called the tensor charge of proton (in general
tensor charge of baryon). The spatial component of $\sigma _{\mu \nu }$ is $%
\Sigma _k=\frac 12\varepsilon _{ijk}\sigma _{ij}$. Use the expansion (5)
again, we obtain 
\begin{eqnarray}
\stackrel{\rightarrow }{\delta q} &=&\int d^3x\overline{\psi }\stackrel{%
\rightarrow }{\Sigma }\psi  \nonumber \\
\ &=&\sum_i\sum_{\lambda \lambda ^{/}}\int d^3k\chi _\lambda ^{\dagger
}\left( \frac m{k_0}\stackrel{\rightarrow }{\sigma }+\frac{\stackrel{%
\rightarrow }{\sigma }\cdot \stackrel{\rightarrow }{k}}{k_0\left(
k_0+m_i\right) }i\stackrel{\rightarrow }{\sigma }\times \stackrel{%
\rightarrow }{k}\right) \chi _{\lambda ^{^{\prime }}}\left( a_{i\stackrel{%
\rightarrow }{k}\lambda }^{\dagger }a_{i\stackrel{\rightarrow }{k}\lambda
^{^{\prime }}}+b_{i\stackrel{\rightarrow }{k}\lambda ^{^{\prime }}}^{\dagger
}b_{i\stackrel{\rightarrow }{k}\lambda }\right)  \nonumber \\
&&\ \ -\sum_i\sum_\lambda \int d^3k\frac{\vec{k}}{k_0}
a_{i\stackrel{\rightarrow }{k}\lambda
}^{\dagger }b_{i-\stackrel{\rightarrow }{k}\lambda }^{\dagger }+h.c.\nonumber\\
&=&\sum_i\sum_{\lambda\lambda^\prime}\int d^3k\chi^{\dagger}_{\lambda}
\left(\stackrel{\rightarrow}{\sigma}-\frac{\stackrel{\rightarrow}{\sigma}
\cdot\stackrel{\rightarrow}{k}\stackrel{\rightarrow}{k}}{k_0\left(k_0+m_i
\right)}\right)\chi_{\lambda^\prime}\left(a^{\dagger}_{ik\lambda}
a_{ik\lambda^\prime}+b^{\dagger}_{ik\lambda^\prime}b_{ik\lambda}\right)
\nonumber\\
&&\ \ -\sum_i\sum_\lambda \int d^3k\frac{\vec{k}}{k_0}
a_{i\stackrel{\rightarrow }{k}\lambda
}^{\dagger }b_{i-\stackrel{\rightarrow }{k}\lambda }^{\dagger }+h.c..
\end{eqnarray}

Under the same approximation as mentioned above for Eq.(4), i.e., neglecting 
the antiquark and the pair creation (annihilation) term and assuming the 
quark momentum distribution of nucleon ground state to be spherically 
symmetric, the second expression of Eq.(17) reduces to the Melosh rotation result discussed by
I. Schmidt and J. Soffer \cite{a20}.

>From Eq.(10), (12)-(15) and (17), we obtain 
\begin{eqnarray}
\vec{\mu _B} &=&\sum_i\frac{Q_i}{m_i^{eff}}\left( 1+\frac{m_i}{m_i^{eff}}\right)
\left( \stackrel{\rightarrow }{S}_i^{NR}-\stackrel{\rightarrow }{S}_{%
\overline{i}}^{NR}\right) +\sum_i\frac{Q_i}{2m_i^{eff}}\left( \stackrel{%
\rightarrow }{L}_i^{NR}-\stackrel{\rightarrow }{L}_{\overline{i}}^{NR}\right)
\ -\sum_i\frac{Q_i}{2m_i^{eff}}\frac 12\stackrel{\rightarrow }{\delta q_i}
\nonumber \\
&&\ -\sum_i\sum_\lambda \frac{Q_i}{2m_i^{eff}}\int d^3k\left( \frac{%
\stackrel{\rightarrow }{k}}{2k_0}+\frac{\stackrel{\rightarrow }{k}}{2\left(
k_0+m_i\right) }\right) a_{i\stackrel{\rightarrow }{k}\lambda }^{\dagger
}b_{i-\stackrel{\rightarrow }{k}\lambda }^{\dagger}+h.c.  \nonumber \\
&&\ -\sum_i\sum_{\lambda \lambda ^{/}}\int d^3k\frac{Q_i}{2k_0}\chi _\lambda
^{\dagger }\left( m_i\stackrel{\rightarrow }{\sigma }+\frac{\stackrel{%
\rightarrow }{\sigma }\cdot \stackrel{\rightarrow }{k}}{k_0+m_i}i\stackrel{%
\rightarrow }{\sigma }\times \stackrel{\rightarrow }{k}\right) \chi
_{\lambda ^{^{\prime }}}\times a_{i\stackrel{\rightarrow }{k}\lambda
}^{\dagger }i\stackrel{\rightarrow }{\partial }_kb_{i-\stackrel{\rightarrow 
}{k}\lambda ^{^{\prime }}}^{\dagger }+h.c.  \nonumber \\
\ &=&\sum_i\frac{Q_i}{2m_i^{eff}}\left( \stackrel{\rightarrow }{L}_i^{NR}-%
\stackrel{\rightarrow }{L}_{\overline{i}}^{NR}\right) +\sum_i\frac{Q_i}{%
m_i^{eff}}\left( 1-\frac{m_i}{2\left( m_i^{eff}+m_i\right) }\right) \left( 
\stackrel{\rightarrow }{S}_i-\stackrel{\rightarrow }{S}_{\overline{i}}\right)
\nonumber \\
&&\ +\sum_i\frac{Q_i}{2m_i^{eff}}\frac{m_i^{eff}}{\left(
m_i^{eff}+m_i\right) }\frac 12\stackrel{\rightarrow }{\delta q_i}  \nonumber
\\
&&\ -\sum_i\sum_{\lambda \lambda ^{/}}\int d^3k\frac{Q_i}{2k_0}\chi _\lambda
^{\dagger }\left( m_i\stackrel{\rightarrow }{\sigma }+\frac{\stackrel{%
\rightarrow }{\sigma }\cdot \stackrel{\rightarrow }{k}}{k_0+m_i}i\stackrel{%
\rightarrow }{\sigma }\times \stackrel{\rightarrow }{k}\right) \chi
_{\lambda ^{^{\prime }}}\times a_{i\stackrel{\rightarrow }{k}\lambda
}^{\dagger }i\stackrel{\rightarrow }{\partial }_kb_{i-\stackrel{\rightarrow 
}{k}\lambda ^{^{\prime }}}^{\dagger }+h.c.
\end{eqnarray}
Neglect the
pair correction terms, we have 
\begin{eqnarray}
\stackrel{\rightarrow }{\mu }_B &=&\sum_i\frac{Q_i}{2m_i^{eff}}\left( 
\stackrel{\rightarrow }{L}_i^{NR}-\stackrel{\rightarrow }{L}_{\overline{i}%
}^{NR}\right)\nonumber\\ &&+\sum_i\frac{Q_i}{m_i^{eff}}\left(\left(1-\frac{m_i}{2
\left(m_i^{eff}+m_i\right)}\right)\left( \stackrel{\rightarrow }{S}%
_i-\stackrel{\rightarrow }{S}_{\overline{i}}\right)+\frac 14
\frac{m_i^{eff}}{m_i^{eff}+m_i}\stackrel{\rightarrow }{\delta q_i}\right)  \nonumber \\
&=&\sum_i\frac{Q_i}{2m_i^{eff}}\left( \stackrel{\rightarrow }{L}_i^{NR}-%
\stackrel{\rightarrow }{L}_{\overline{i}}^{NR}\right) +\sum_i\frac{Q_i}{%
m_i^{eff}}\left(\left(1+\frac{m_i}{m_i^{eff}}\right)\left( 
\stackrel{\rightarrow }{S}_i^{NR}-\stackrel{\rightarrow }{S}%
_{\overline{i}}^{NR}\right)-\frac 14\stackrel{\rightarrow }{\delta q_i}\right) .
\end{eqnarray}
For ground state baryons, it is plausible to assume 
\begin{equation}
\left\langle B\left| \stackrel{\rightarrow }{L}_i^{NR}\right| B\right\rangle
=\left\langle B\left| \stackrel{\rightarrow }{L}_{\overline{i}}^{NR}\right|
B\right\rangle =0.
\end{equation}
Then we have 
\begin{eqnarray}
\mu _B &=&\sum_i\frac{Q_i}{2m_i^{eff}}\left(\left(1+\frac{m_i}{m_i^{eff}}\right)\left( \Delta _i^{NR}-\Delta _{%
\overline{i}}^{NR}\right)-\frac 12\delta q_i\right)  \nonumber \\
\ &=&\sum_i\frac{Q_i}{2m_i^{eff}}\left(\left(1-\frac{m_i}{2
\left(m_i^{eff}+m_i\right)}\right)\left( \Delta _i-\Delta _{\overline{i}}\right)+%
\frac 12\frac{m_i^{eff}}{m_i^{eff}+m_i}\delta q_i\right)
\end{eqnarray}
where 
\begin{eqnarray}
\Delta _i &=&2\left\langle B\left| S_{i3}\right| B\right\rangle  \nonumber \\
\Delta _{\overline{i}} &=&2\left\langle B\left| S_{\overline{i}3}\right|
B\right\rangle  \nonumber \\
\Delta _i^{NR} &=&2\left\langle B\left| S_{i3}^{NR}\right| B\right\rangle 
\nonumber \\
\Delta _{\overline{i}}^{NR} &=&2\left\langle B\left| S_{\overline{i}%
3}^{NR}\right| B\right\rangle
\end{eqnarray}
Therefore the Karl-Sehgal relations of the octet baryon magnetic moments
upgraded by Cheng and Li \cite{a21} should be upgraded further 
\begin{eqnarray}
\mu _P &=&\frac{2e}{3m_u}W_u+\frac{-e}{%
3m_d}W_d+\frac{-e}{3
m_s}W_s,  \nonumber \\
\mu _n &=&\frac{-e}{3m_d}W_u+\frac{2e}{%
3m_u}W_d+\frac{-e}{
m_s}W_s,  \nonumber \\
\mu _{\Sigma ^{+}} &=&\frac{2e}{3m_u}W_u+\frac{%
-e}{3m_s}W_d+\frac{-e}{3
m_d}W_s,  \nonumber \\
\mu _{\Sigma ^{-}} &=&\frac{-e}{3m_d}W_u+\frac{%
-e}{3m_s}W_d+\frac{2e}{3
m_u}W_s,  \nonumber \\
\mu _{\Xi ^0} &=&\frac{-e}{3m_s}W_u+\frac{2e}{%
3m_u}W_d+\frac{-e}{3
m_d}W_s,  \nonumber \\
\mu _{\Xi ^{-}} &=&\frac{-e}{3m_s}W_u+\frac{-e}{%
3m_d}W_d+\frac{2e}{3
m_u}W_s,  \nonumber \\
\mu _\Lambda &=&\frac 16\left( \frac{2e}{3m_u}+%
\frac{-e}{m_d}\right) \left(
W_u+4W_d+W_s\right)  \nonumber \\
&&\ \ \ +\frac 16\cdot \frac{-e}{3m_s}\left(
4W_u-2W_d+4W_s\right) ,  \nonumber \\
\mu _{\Lambda \Sigma ^0} &=&-\frac 1{2\sqrt{3}}\left( \frac{2e}{%
3m_u}-\frac{-e}{3
m_d}\right) \left( W_u-2W_d+W_s\right) . 
\end{eqnarray}
where 
\begin{eqnarray}
W_i&=&\frac 12\left(\left(1+\frac{m_i}{m_i^{eff}}\right)\left(
 \Delta _i^{NR}-\Delta _{\overline{i}}^{NR}\right)-\frac 12\delta
q_i\right)\nonumber\\
&=&\frac{1}{2}\left(\left(1-\frac{m_i}{2
\left(m_i^{eff}+m_i\right)}\right)\left(\Delta_i-\Delta_{\bar{i}}\right)
+\frac{1}{2}\frac{m_i^{eff}}{m_i^{eff}+m_i}\delta q_i\right)
\end{eqnarray}
and the effective quark mass $m_i^{eff}$ has been written directly by
constituent quark mass $m_u$, $m_d$ and $m_s$. it should be noted that these
relations are now based on QCD but with the following approximation 
\begin{eqnarray}
&&k_0\text{ approximated by }\left\langle k_0\right\rangle =m_i^{eff}, 
\nonumber \\
&&{\rm Pair~creation~terms~}{\rm neglected.}
\end{eqnarray}

An approximated relation between the nonrelativistic spin $\stackrel{
\rightarrow}{S}_i^{NR}$, axial
vector current and tensor current operators can be obtained as well.
>From Eq.(10), (14), and (17), neglecting the pair creation and annihilation
terms we have 
\begin{equation}
\stackrel{\rightarrow }{S}_q^{NR}=\sum_i\frac 1{1+\frac{m_i}{\left\langle
k_0\right\rangle }}\left( \stackrel{\rightarrow }{S}_i+\stackrel{\rightarrow 
}{S}_{\overline{i}}+\frac 12\stackrel{\rightarrow }{\delta q_i}+\stackrel{%
\rightarrow }{\delta q_{\overline{i}}}\right)
\end{equation}
where 
\begin{equation}
\stackrel{\rightarrow }{\delta q_{\overline{i}}}=-\sum_{\lambda \lambda
^{/}}\int d^3k\chi _\lambda ^{\dagger }\left( \frac m{k_0}\stackrel{%
\rightarrow }{\sigma }+\frac{\stackrel{\rightarrow }{\sigma }\cdot \stackrel{%
\rightarrow }{k}}{k_0\left( k_0+m_i\right) }i\stackrel{\rightarrow }{\sigma }%
\times \stackrel{\rightarrow }{k}\right) \chi _{\lambda ^{^{\prime }}}b_{i%
\stackrel{\rightarrow }{k}\lambda ^{^{\prime }}}^{\dagger }b_{i\stackrel{%
\rightarrow }{k}\lambda }
\end{equation}

Under further approximations mentioned for Eq.(4) and (17), i.e., in the 
pure valence quark and spherically 
symmetric momentum distribution approximation,
by means of the second expression of $\stackrel{\rightarrow }{\delta q}$ in Eq.(17), one can
obtain another 
relation between nonrelativistic spin sum $\Delta \Sigma$, axial charge
$\Delta q$ and tensor charge $\delta q$ of proton discussed by I. Schmidt,
J. Soffer, Ma and He\cite{a22} which reads as $\Delta \Sigma +
\Delta q= 2\delta q$.

\section{A valence-sea mixing constituent quark model calculation}

Constituent quark model is the most successful one in low energy hadron
physics. Even for the polarization asymmetry in the valence region in DIS,
the constituent quark model still gave a historical successful prediction 
\cite{a14}. The OZI rule violation and the nucleon spin structure studies do
remind us that pure valence configuration is an approximation and the sea
quark components should be taken into account. From a general view point of
Fock space expansion, a baryon should be described by 
\begin{equation}
\left| B\right\rangle _\alpha =C_0\left| q^3\right\rangle _\alpha +\sum
C_i\left| q^3q\overline{q}\right\rangle _{i\alpha }+\cdot \cdot \cdot
\end{equation}
where the higher Fock components have been omitted. Chiral quark model leads
to a similar description of baryons, where $q\overline{q}$ is replaced by a
Goldstone boson (pseudo-scalar meson) \cite{a23}. Because we can't do a
nonpertubative QCD calculation to check what we have discussed so far, a
valence-sea quark mixing constituent quark model calculation has been done
under these inspirations \cite{a24}. The model Hilbert space is assumed to
be consisted of pure valence component $q^3$ and all possible combinations
compatible with the quantum number of a baryon with colorless $s$-wave
octet and decuplet $q^3$ combined with $q\overline{q}$ having pseudo-scalar
quantum numbers. To meet the positive parity condition of the ground octet
and decuplet baryons, the relative motion between $q^3$ and $q\overline{q}$
centers is assumed to be in $p$-wave. The internal wavefunction of $q^3$ and 
$q\overline{q}$ and the relative motion wavefunction are all assumed to be a
Gaussian one with a common size parameter $b$ for simplicity of the
numerical calculation. The model Hamiltonian is almost the same as those of
the Isgur model\cite{a9} except that a $q\overline{q}$ pair creation and annihilation
interaction term has been included to mix the $q^3$ and $q^3q\overline{q}$
components. 
\begin{eqnarray}
\ &&H=\sum_i\left(m_i+{\frac{p_i^2}{2m_i}}\right)+\sum_{i<j}%
\left(V_{ij}^c+V_{ij}^G\right)+%
\sum_{i<j}\left(V_{i,i^{\prime }j^{\prime }j}+V_{i,i^{\prime }j^{\prime }j}^{\dag
}\right),  \nonumber \\
\ &&V_{ij}^c=-a_c\vec \lambda _i\cdot \vec \lambda _jr_{ij}^2,  \nonumber \\
\ &&V_{ij}^{Gs}=\alpha _s{\frac{\vec \lambda _i\cdot \vec \lambda _j}4}\left({%
\frac 1{r_{ij}}}-{\frac \pi 2}\left({\frac 1{m_i^2}}+{\frac 1{m_j^2}}+{\frac{2%
\vec \sigma _i\cdot \vec \sigma _j}{3m_im_j}}\right)\delta \left(\vec r_{ij}\right)+\cdot
\cdot \cdot \right),  \nonumber \\
\ &&V_{ij}^{Ga}=\alpha _s\left({\frac{\vec \lambda _i\cdot \vec \lambda _j^{*}}2}%
\right)^2\left({\frac 13}+{\frac{\vec f_i\cdot \vec f_j^{*}}2}\right)\left({\frac{\vec \sigma
_i\cdot \vec \sigma _j}2}\right)^2{\frac 23}{\frac 1{(m_i+m_j)^2}}\delta \left(\vec r%
_{ij}\right),  \nonumber \\
\ &&V_{i,i^{\prime }j^{\prime }j}=i\alpha _s{\frac{\vec \lambda _i\cdot \vec 
\lambda _j}4}{\frac 1{2r_{ij}}}\left(\left(\left({\frac 1{m_i}}+{\frac 1{m_j}}\right)\vec \sigma
_j+{\frac{i\vec {\sigma _j}\times \vec {\sigma _i}}{m_i}}\right)\cdot {\frac{\vec r%
_{ij}}{r_{ij}^2}}-{\frac{2\vec \sigma _j\cdot \vec \nabla _i}{m_i}}\right),
\end{eqnarray}
where $\vec \lambda _i\left(\vec f_i\right)$ are the $SU_3^c\left(SU_3^f\right)$ Gell-Mann
operators, the $V_{ij}^{Gs}$, $V_{ij}^{Ga}$ and $V_{i,i^{\prime }j^{\prime
}j}$ correspond to the following diagrams of Fig.1 respectively, i.e., we
use an effective one gluon exchange to derive the quark interactions except
the confinement part which is introduced phenomenologically. The other
symbols have their usual meaning.

\fbox{fig.1 goes here.}

The model parameters, $u,d$ quark mass $m$, $s$ quark mass $m_s$, quark
gluon coupling constant $\alpha _s$, $q^3$ quark core baryon size $b$, and
confinement strength $a_c$, are fixed by an overall fit to the ground state
octet and decuplet baryon properties.

\fbox{Table I goes here.}

\fbox{Table II goes here.}

Table I shows the wave function of the proton obtained from the
diagonalization of the Hamiltonian (29) in Fock space (28). The entry is the
amplitude of the individual component. The total sea quark component is
about $15\%$. It is an example of our model wave functions of ground state
baryons.

Table II summarize our model predictions and the model parameters. These
results show that it is possible to have a valence and sea quark mixing
model which can describe, with the commonly accepted quark model parameters,
the ground state octet and decuplet baryon properties as good as the
successful pure valence quark model. Furthermore, the proton charge radius
is reproduced as well. A too small proton charge radius has been a long
standing problem of the constituent quark model. The first excited states
are higher than 2 GeV. This is consistent with the fact that there is no
pentaquark states observed below 2 GeV.

\fbox{Table III goes here.}

The spin content of the proton is listed in table III, where the matrix
element of the axial vector current operator (4) in a spin up proton state
is decomposed into particle number conserved components $q^3\leftrightarrow
q^3,q^4\bar q\leftrightarrow q^4\bar q$ and particle number nonconserved
components $q^3\leftrightarrow q^4\bar q$. In doing this calculation, the
quark (antiquark) operator $a_{iks}^{\dagger }$ ($b_{iks}^{\dagger }$) in
Eq.(4) has been identified with the constituent quark (antiquark) degree of
freedom. This is a model assumption and is usual for quark model
calculations. The second column lists the axial charge of the pure valence
configuration $q^3$, there the relativistic correction (second term in
Eq.(4)) and the normalization factor $\left( -0.923\right) ^2$ have been
included. The sum $\Delta u+\Delta d+\Delta s=0.773-0.193+0=0.580$ of the $%
q^3$ configuration divided by the normalization factor gives $0.580/\left(
-0.923\right) ^2=0.681$ which shows even for a pure $q^3$ configuration the
axial charge $\Delta q$ is already different from the spin sum $\Delta
\Sigma =1$ due to the relativistic correction. The fourth column lists the
axial charge of the $q^4\bar q$ configuration. The sea quark contribution
can't be separated due to quark antisymmetrization. Antiquark contribution
(which has not been listed in table III) is quite small. This is the same 
as the chiral quark model result obtained by Cheng and Li\cite{a23}.
The main reduction
is due to $q^3\leftrightarrow q^4\bar q$ transition term. Physically it is
similar to the generalized Sullivan processes discussed by Hwang {\it et al} 
\cite{a16} and makes our model different from the chiral quark model\cite{a23}. 
The sum of the three terms listed in column 5 is quite close to
the world average value $\Delta q$ listed in column 6 and the lattice QCD
results listed in column 7\cite{a25,a26}.

Tensor charge has been calculated by the operator in Eq.(17). It has the
relativistic correction and pair creation (annihilation) correction too.
Again, it is reduced due to relativistic correction and pair creation
 and annihilation term,
but
the effect is small for the tensor charge in comparison with those for the
axial charge. Up to now tensor charge has not been measured. Fortunately a
lattice QCD calculation has just been published \cite{a27}. Even though
their axial charge $\Delta q$ does not match the experimental results as well
as that of Dong, Lagae, and Liu \cite{a26}, it might still show our model
tensor charge is close to the reality\cite{a27}.

Even though the spin structure of barons is complicated than the naive
pure valence quark model, the magnetic moments of baryons are fitted as 
well as the naive ones.

These results show that the nucleon spin structure information obtained in
polarized DIS is possible to be described in a constituent quark model. Of
course, the sea quark components should be taken into account. However, to
fit the DIS measured nucleon spin structure, only about $15\%$ sea quark
component is needed. This means the naive pure valence quark model is a
reasonable approximation. We would like to emphasize that we are certainly
not pretend to claim having a good nucleon model. On the contrary, our model
is a rough one, many points should be improved, and many points should be
checked. The aim to show this model results is to pass a message that
constituent quark model, after including small amount ($\sim 15\%$) sea
quark component, is compatible with the nucleon spin structure discovered in
DIS.

\section{Discussions and conclusions}

Deep inelastic scattering detects the inner structure of nucleon directly and
played vital role in establishing the quark model of hadrons. Constituent 
quark model is the most successful one in explaining the hadron properties.
Polarized lepton-nucleon deep inelastic scattering reveals the spin structure
of nucleon. 
Whether or not nucleon spin structure is consistent with the constituent 
quark model is controversial still.

The main messages of this report are:
\begin{enumerate}
\item In a truncated Fock space of a baryon with only effective quark degree 
of freedom, baryon spin operator can be decomposed in two equivalent ways
\begin{equation}
\vec{J}=\vec{S}^{NR}+\vec{L}^{NR}=\vec{S}+\vec{L}
\end{equation}
$\vec{S}^{NR}$ and $\vec{L}^{NR}$ are the quark spin and orbital angular 
momentum operators used in the nonrelativistic constituent quark model, 
while $\vec{S}$ and $\vec{L}$ are the quark spin (i.e., axial vector current
operator) and orbital angular momentum operators derived from QCD Lagrangian.
\item The axial charge $\Delta q$ extracted from the polarized lepton-proton
deep inelastic scattering is related to the matrix element of $\vec{S}$ in a
polarized proton and should not be identified to the constituent quark
model spin sum $\Delta\Sigma$ which is related to the matrix element of 
$\vec{S}^{NR}$.
\item Constituent quark model, after mixing a small amount(15\%) of sea 
quark components, is able to describe the fact that $\Delta q \sim \frac 13
\Delta\Sigma$. The reduction of $\Delta q$ is due to relativistic correction
and the transition matrix element $\left\langle q^3\left |\vec{S}\right |
q^3q\bar{q}\right\rangle$.
\item The successful relation between magnetic moments and spins of baryons 
first obtained from constituent quark model are a robust relation. Even
though it should be upgraded, the main feature of the Sehgal-Karl-Cheng-Li
relations and the good fitting of baryon magnetic moments remain there even
though the nucleon spin structure is complicated as revealed in polarized DIS. 
\item The quantitative fit of baryon properties reported here, especially 
the axial charge $\Delta q$ and tensor charge $\delta q$ of proton, is model
dependent. The main messages mentioned above are not a model one except the
numerical values of $\left\langle q^3\left |\vec{S}\right |
q^3q\bar{q}\right\rangle$. This transition matrix element contribution
to the axial charge $\Delta q$ might be a quark model version of the gluon
contribution discussed in perturbative QCD calculation. Of course this point
is tentative and further studies are needed.
\end{enumerate}

The nucleon structure is certainly more complicated than the naive 
constituent quark model. However constituent quark model is a good 
approximation. The nucleon spin structure discovered in polarized deep 
inelastic scattering invites the improvement of constituent quark model,
but does not invalidate it.

This work is supported in part by the NSF (19675018), SEDC and SSTC of China.
Helpful discussion with B.Q. Ma and H.X. He are acknowledged.

\pagebreak
\begin{center}
FIGURE CAPTION
\vspace{3cm}

Fig.1 quark interaction diagrams
\end{center}
\begin{table}
\begin{center}
\caption{proton model wave function}
\begin{tabular}{|c|c|c|c|c|c|c|c|}
$q^3$&$N\eta$&$N\pi$&$\Delta\pi$&$N\eta^\prime$&$\Lambda K$&$\Sigma K$
&$\Sigma^* K$ \\ \hline
$-$0.923&0.044&0.232&$-$0.252&0.065&0.109&$-$0.036&$-$0.106 \\
\end{tabular}
\end{center}
\end{table}

\begin{table}
\begin{center}
\caption{masses and magnetic moments of the baryon
octect and decuplet.
$m=330(MeV), m_s=564(MeV), b=0.61(fm), \alpha_s=1.46, a_c=48.2 (MeVfm^{-2})$}
\begin{scriptsize}
\begin{tabular}{|c|c|c|c|c|c|c|c|c|c|c|c|c|}
    &   &p  &n   &$\Lambda$&$\Sigma^+$&$\Sigma^-$&$\Xi^0$&$\Xi^-$&
 $\Delta$&$\Sigma^*$&$\Xi^*$&$\Omega$ \\ \hline
    &M(Mev)&\multicolumn{2}{c|}{939}&1116&\multicolumn{2}{c|}{1193}
 &\multicolumn{2}{c|}{1346}&1232&1370&1523&1659 \\ \cline{2-13}
Theor.&E1(MeV)&\multicolumn{2}{c|}{2203}&2323&\multicolumn{2}{c|}{2306}
&\multicolumn{2}{c|}{2409}&2288&2306&2450&2638 \\ \cline{2-13}
    &$\mu(\mu_N)$&2.780&$-$1.818&$-$0.522&2.652&$-$1.072&$-$1.300&$-$0.412
& & & & \\
\cline{2-13}
    &$\sqrt{\langle r^2 \rangle}(fm)$&0.802&0.124& & & & & & & & & \\ \hline
    &M(MeV)&\multicolumn{2}{c|}{939}&1116&\multicolumn{2}{c|}{1189}
&\multicolumn{2}{c|}{1315}&1232&1385&1530&1672 \\ \cline{2-13}
Exp.&$\mu(\mu_N)$&2.793&$-$1.913&$-$0.613&2.458&$-$1.160&$-$1.250&$-$0.651
& & & & \\
\cline{2-13}
    &$\sqrt{\langle r^2 \rangle}(fm)$&0.836&0.34& & & & & & & & & \\
\end{tabular}
\end{scriptsize}
\end{center}
\end{table}

\begin{table}
\begin{center}
\caption{The spin content and tensor charge of proton}
\begin{tabular}{|c|c|c|c|c|c|c|c|}
  &$q^3$&$q^3-q^4\bar{q}$&$q^4\bar{q}-q^4\bar{q}$&sum&exp.&lattice\cite{a26}&lattice\cite{a27} \\ \hline
$\Delta u$&0.773&$-$0.125&0.100&0.75&0.81&0.79(11) &0.638(54)  \\ \hline
$\Delta d$&$-$0.193&$-$0.249&$-$0.041&$-$0.48&$-$0.44&$-$0.42(11) &$-$0.347(46)\\ \hline
$\Delta s$&0    &$-$0.064&$-$0.002&$-$0.07&$-$0.10&$-$0.12(1) &$-$0.109(30)\\ \hline
$\delta u$&0.955&$-$0.123&0.127&0.959& & & 0.839(60)\\\hline
$\delta d$&$-$0.239&$-$0.061&$-$0.047&$-$0.347& & &$-$0.231(55)\\ \hline
$\delta s$&0    &$-$0.022&$-$0.002&$-$0.024& & &$-$0.046(34)\\
\end{tabular}
\end{center}
\end{table}


\begin{references}
\bibitem{a1}  E. Fermi and C.N. Yang, Phys. Rev. {\bf 76}, 1739 (1949).

\bibitem{a2}  S. Sakata, Prog. Theor. Phys. {\bf 16}, 686 (1956).

\bibitem{a3}  M. Gell-Mann, Phys. Lett {\bf 8}, 214 (1964).

\bibitem{a4}  G. Zweig, CERN Reports, {\bf TH401} and {\bf 412} (1964).

\bibitem{a5}  R.E. Taylor, Rev. Mod. Phys. {\bf 63}, 573 (1991); H.W. Kendall,
{\it ibid}, 597; J.I. Friedman, {\it ibid}, 615 and references there in.

\bibitem{a6}  R.P. Feynman, Phys. Rev. Lett. {\bf 23}, 1415 (1969); {\it Photon 
Hadron Interactions} (W.A. Benjamin, New York, 1972); J.D. Bjorken and E.A. 
Paschos, Phys. Rev. {\bf 185}, 1975 (1969).

\bibitem{a7}  G. 't Hooft, Proc. Colloquium on Renormalization of Yang-Mills
Fields and Application to Particle Physics, Marseilles, 1972 (ed. C.P. 
Korthals-Altes); D.J. Gross and F. Wilczek, Phys. Rev. Lett. {\bf 30},
1343 (1973); H.D. Politzer, Phys. Rev. Lett. {\bf 30}, 1346 (1973); 
H. Fritzsch, M. Gell-Mann and H. Leutwyler, Phys. Lett. {\bf 46B}, 365 (1973).

\bibitem{a8}  A. de Rujula, H. Georgi, and S.L. Glashow, Phys. Rev. D {\bf 12},
147 (1975); A. Chodos {\it et al.}, Phys. Rev. D{\bf 9}, 3471 (1974); 
D {\bf 10}, 2599 (1974); 
T. De Grand {\it et al.}, Phys. Rev. D {\bf 12}, 2060 (1975); R. Friedberg
and T. D. Lee, Phys. Rev. D {\bf 16}, 1096
(1977); D {\bf 18}, 2623 (1978); T.H.R. Skyrme, Proc. Roy. Soc. {\bf 260}, 
127 (1961); Nucl. Phys. {\bf 31}, 556 (1962); E. Witten, Nucl. Phys.
{\bf B223}, 422, 433 (1983).

\bibitem{a9}  N. Isgure and G. Karl, Phys. Rev. D {\bf 18}, 4187 (1978);
D {\bf 19}, 2653 (1979); D {\bf 20}, 1191 (1979).

\bibitem{a10}  F. Wang {\it et al.}, Proceedings of 1978 National Conference
of Physics, (Atomic Energy Pub. Co., Beijing, 1979) P.68 (in Chinese); 
Contributed Papers Book of Workshop on Nucl. Phys. with Real and Virtual
Photons, Bologna, Italy, Nov. 25-28, 1980, P.2-4; F. Wang {\it et al.},
Phys. Rev. Lett. {\bf 69}, 2901 (1992); G.H. Wu {\it et al.}, Phys. Rev.
C {\bf 53}, 1161 (1995); M. Oka and K. Yazaki, Phys Lett. {\bf 90B}, 41
(1980); Prog. Theor. Phys. {\bf 66}, 556, 572 (1981);
S. Takeuchi, K. Shimizu and K. Yazaki, Nucl. Phys.
{\bf A504}, 777 (1989); A. Faessler {\it et al.}, Phys. Lett. {\bf 112B},
201 (1982); Nucl. Phys. {\bf A402}, 555 (1983); {\bf A578}, 573 (1994);
Y. Fujwara and K.T. Hencht, Nucl. Phys. {\bf A444}, 541 (1985); Y. Fujiwara
{\it et al.}, Phys. Rev. Lett. {\bf 76}, 2242 (1996).

\bibitem{a11}  S. Weinberg, Phys. Rev. Lett. {\bf 65}, 1181 (1990).

\bibitem{a12}  J. Ashman {\it et al.}, Phys. Lett. {\bf 206B}, 364 (1988);
Nucl. Phys. {\bf B328}, 1 (1989)

\bibitem{a13}  R.L. Jaffe, Phys. Today, {\bf 48}(9), 24 (1995); M. Anselmino,
A. Efremov and E. Leader, Phys. Rep. {\bf 261}, 1 (1995); V. Stiegler, Phys.
Rep. {\bf 277}, 1 (1996); H.Y. Chen, Int. J. Mod. Phys. A {\bf 11}, 5109 (1996);
G.M. Shore, Report No. hep-ph/9710367.

\bibitem{a14}  F.E. Close and R.G. Roberts, Phys. Lett. {\bf 316B}, 165 (1993);
{\bf 336B}, 257 (1994); F.E. Close, Report No. hep-ph/9509251.

\bibitem{a15}  B.Q. Ma, J. Phys. G {\bf 17}, L53 (1991); B.Q. Ma and 
Q.R. Zhang, Z. Phys. C {\bf 58}, 479 (1993).

\bibitem{a16}  W-Y. P. Hwang, J. Speth and G. E. Brown, Z. Phys. A {\bf 339},
383 (1991); C.H. Chung and W-Y. P. Hwang, Phys. Rev. D {\bf 49}, 2221 (1994).

\bibitem{a17}  T. Ellis and M. Karliner, Report. No. hep-ph/9601280, 
CERN-TH/95-334.

\bibitem{a18}  R.L. Jaffe and A. Manohar, Nucl. Phys. {\bf B337}, 509 (1990); 
X. Ji, Phys. Rev. Lett. {\bf 78}, 610 (1997); X.S. Chen and 
F. Wang, Commun. Theor. Phys. {\bf 27}, 121 (1997).

\bibitem{a19}  R.L. Jaffe and X. Ji, Phys. Rev. Lett. {\bf 67}, 552 (1991) and 
references there in.

\bibitem{a20}  I. Schmidt and J. Soffer, Phys. Lett. {\bf 407B}, 331 (1997).

\bibitem{a21}  G. Karl, Phys. Rev. D {\bf 45}, 247 (1992); T. P. Cheng and 
L.F. Li, Phys. Lett. {\bf 366B}, 365 (1996).

\bibitem{a22}  B.Q. Ma, I. Schmidt and J. Soffe, Preprint CPT-97/3539, 
BIHEP-TH-97-13, USM-TH-68; H.X. He, CIAE preprint; and private 
communications.

\bibitem{a23}  S. Weinberg, Physica A {\bf 96}, 327 (1979); A. Manohar and 
H. Georgi, Nucl. Phys. {\bf B234}, 189 (1984); E.J. Eichten, I. Hinchliffe 
and C. Quigg, Phys. Rev. D {\bf 45}, 2269 (1992); T.P. Cheng and L.F. Li, 
Phys. Rev. Lett {\bf 74}, 2872 (1995); X. Song, J.S. McCarthy and H.J. Weber,
Phys. Rev. D {\bf 55}, 2624 (1997).

\bibitem{a24} A preliminary result has been published in Phys. Rev. C {\bf 57}, 
R31 (1998).

\bibitem{a25}  M. Fukugita {\it et al.}, Phys. Rev. Lett. {\bf 75}, 2092 (1995).

\bibitem{a26}  S.J. Dong, J.P. Lagae and K.F. Liu, Phys. Rev. Lett. {\bf 75},
2096 (1995).

\bibitem{a27}  S. Aoki {\it et al.}, Phys. Rev. D {\bf 56}, 433 (1997).

\bibitem{a28}  B.L. Ioffe and A. Khodjamiraian, Phys. Rev. D {\bf 51}, 3373 
(1995); H.X. He and X. Ji, Phys. Rev. D {\bf 52}, 2690 (1995); H.C. Kim, 
M. Polyakov and K. Goeke, Phys. Rev. D {\bf 53}, R4715 (1996); Phys. Lett. 
{\bf 387B}, 577 (1996).
\end{references}
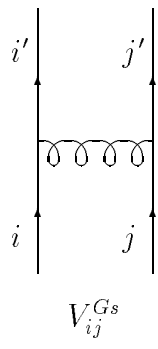
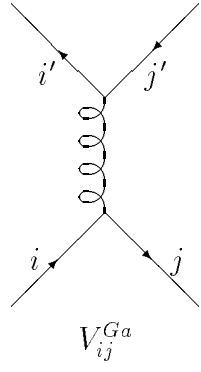
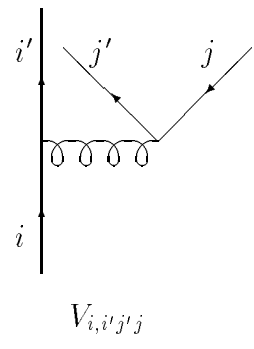
\end{document}